\documentclass[useAMS,usenatbib]{mn2e}

\usepackage[T1]{fontenc}
 \usepackage{pslatex}
\usepackage{amsmath}
\usepackage{amsfonts}
\usepackage{color}
\usepackage{url}
\usepackage{bm}
\usepackage{graphicx}
\usepackage{subfigure}
\usepackage{amssymb}
\usepackage{soul}
\usepackage{booktabs}
\usepackage{array}
\usepackage[utf8]{inputenc}
\inputencoding{latin1}
\inputencoding{utf8}

\def\lsim{ \lower .75ex\hbox{$\sim$} \llap{\raise .27ex \hbox{$<$}} }
\def\gsim{ \lower .75ex \hbox{$\sim$} \llap{\raise .27ex \hbox{$>$}} }

\title[Anisotropic electron populations in BL Lacs] 
{Anisotropic electron populations in BL Lac jets: consequences for the observed emission}

\author[Tavecchio \& Sobacchi]
{F. Tavecchio$^1$\thanks{E--mail: fabrizio.tavecchio@inaf.it}, E. Sobacchi$^{2,3}$\\
$^1$INAF -- Osservatorio Astronomico di Brera, via E. Bianchi 46, I--23807 Merate, Italy\\
$^2$Physics Department, Ben-Gurion University, P.O.B. 653, Beer-Sheva 84105, Israel\\
$^3$Department of Natural Sciences, The Open University of Israel, P.O.B. 808, Raanana 4353701, Israel\\
}

\begin{document}

% \date{Accepted 1988 December 15. Received 1988 December 14; 
% in original form 1988 October 11}

%\pagerange{\pageref{firstpage}--\pageref{lastpage}} \pubyear{2007}

\maketitle

\begin{abstract} 
We investigate the impact on the properties of high-energy emitting BL Lac objects of a population of electrons with an anisotropic momentum distribution. We adopt a simple phenomenological description of the anisotropy, in which the most energetic electrons have a small pitch angle and the least energetic electrons are isotropic, as proposed by \citet{SobacchiLyubarsky2019}. We explore (i) a simple model that assumes a phenomenological shape for the electron energy distribution, and (ii) a self-consistent scheme in which the electrons follow a distribution which is the result of the balance between injection and radiative losses (we include the effects of the anisotropy on the synchrotron cooling rate). Considering the BL Lac object Mkn 421 as representative of the entire class, we show that in both cases the emission can be satisfactorily reproduced under equipartition between the magnetic field and the relativistic electrons. This is in better agreement with the idea that jets are launched as Poynting dominated flows with respect to the standard isotropic scenario, which requires both a low magnetization and a low radiative efficiency to reproduce the observed emission. The hard spectrum predicted for the inverse Compton continuum at TeV energies could be used as a potential test of the anisotropic model.
\end{abstract}


\section{Introduction}

Blazars are Active Galactic Nuclei with a relativistic jet pointed towards the Earth \citep[e.g.][]{UrryPadovani1995}. Since the radiation from the jet is strongly beamed (and often completely outshines the other components, such as the accretion flow), blazars represent ideal natural laboratories to test our theoretical understanding of the physical processes acting in relativistic jets.

According to the standard view \citep[e.g.][]{BlandfordZnajek1977}, extragalactic jets are powered by the rotational energy of a supermassive black hole, which is transferred to the outflowing plasma via electromagnetic stresses. These jets start as highly magnetized outflows accelerating through the conversion of the Poynting flux into the kinetic flux under differential collimation. This scheme naturally predicts that the plasma remains highly magnetized, or at most reaches equipartition between the magnetic and the kinetic energy densities, until some dissipative process occurs \citep[e.g.][]{Komissarov2009, Lyubarski2010}.

The development of MHD instabilities may trigger the dissipation of the magnetic energy into the particle kinetic energy at the emission sites, possibly through relativistic magnetic reconnection \citep[e.g.][]{BrombergTchekhovskoy2016, Barniol2017}. Kinetic simulations have shown that relativistic magnetic reconnection is able to accelerate a population of non-thermal particles, whose energy is distributed according to a power law \citep[e.g.][]{SironiSpitkovsky2014, Guo+2014, Werner+2017, Petropoulou+2019}. Interestingly, the power law index found by these studies is not universal, being it dependent on the magnetization of the plasma, on the strength of the guide field, and on the composition of the plasma, which one may expect to be different in different objects.

The Spectral Energy Distribution (SED) of blazars is characterised by two broad non-thermal components, the first one peaking at IR-optical-UV frequencies, and the second one peaking in the $\gamma$ rays. The SED follows a sequence \citep[e.g.][]{Fossati1998, Ghisellini+2017}: the brightest objects, which usually show strong emission lines and are therefore classified as Flat Spectrum Radio Quasars (FSRQ), peak at lower frequencies, while the less powerful objects, which usually show weak or no emission lines and are therefore classified as BL Lacs, peak at higher frequencies. The first bump of the SED is due to the synchrotron radiation from a population of non-thermal electrons, while the second bump is usually attributed to the Comptonization of either an external photon field (in the case of FSRQ), or of the synchrotron photons themselves (in the case of BL Lacs) \citep[e.g.][]{Sikora1994, Sikora2009, Ghisellini1998, Ghisellini2010, Tavecchio1998}.

The classical one-zone Synchrotron-Self-Compton (SSC) model used to reproduce the SED of the BL Lacs challenges our theoretical understanding of relativistic jets in several aspects (\citealt{TavecchioGhisellini2016}; see also \citealt{InoueTanaka2016, NalewajkoGupta2017, Costamante+2018}):
\begin{enumerate}
\item The magnetic energy density is much smaller than the kinetic energy density of the electrons. This is in contrast with the theoretical expectation that the emission sites are highly magnetized.
\item The average radiative efficiency of the electrons (defined as the ratio between the dynamical timescale and the radiative cooling time) is low. However, fitting the SED requires the electrons below a characteristic Lorentz factor $\gamma_{\rm b}$ to be distributed according to a universal power law $N(\gamma)\propto\gamma^{-2}$ \citep[e.g.][]{Tavecchio2010}, which is exactly what one would expect in the fast cooling regime.\footnote{If the cooling is fast, the number of electrons at the energy $\gamma$, namely $\gamma\, N(\gamma)$ where $N(\gamma)$ is the number of electrons per unit energy (or Lorentz factor), is proportional to the cooling time at that energy, $t_{\rm cool}(\gamma)$. Since $t_{\rm cool}\propto\gamma^{-1}$ for both the synchrotron and the inverse Compton processes, one eventually finds that $N(\gamma) \propto \gamma^{-2}$.} Since reconnection is unlikely to produce such a universal energy distribution, the origin of the distribution remains unclear if the radiative losses are inefficient.
\end{enumerate}

These conclusions stem from the fact that in the classical one-zone SSC model the maxima of the two bumps of the SED are produced by the same electrons, whose energy can be robustly derived. A possible way-out to these conclusions is proposed by \citet[][]{SobacchiLyubarsky2019}, who have argued that (i) the non-thermal electrons may be accelerated in the direction of the magnetic field; (ii) the gyro-resonant pitch angle scattering makes the distribution of the electrons isotropic below a critical $\gamma_{\rm iso}$ (of the order of the proton to electron mass ratio), while the pitch angle of the most energetic electrons remains small. If $\gamma_{\rm iso}$ is smaller than the break $\gamma_{\rm b}$ of the electron energy distribution, the two bumps of the SED are produced by different populations of electrons, the synchrotron (inverse Compton) bump being produced by the electrons at $\gamma_{\rm iso}$ ($\gamma_{\rm b}$). As discussed by \citet[][]{SobacchiLyubarsky2019}, in such a scenario the magnetization of the emission sites and the radiative efficiency of the non-thermal electrons required to reproduce the SED may be significantly higher with respect those inferred using the classical one-zone SSC model\footnote{An alternative possibility is that the emission comes from one single reconnecting current sheet, which then fragments into plasmoids of different sizes. \citet[][]{Christie+2019} suggested that the IC emission is boosted as an effect of the relative velocity of different plasmoids, which alleviates the tension on the magnetisation required to reproduce the SED}.

In this paper we develop a quantitative model for the SED of blazars based on the anisotropy proposed by \citet[][]{SobacchiLyubarsky2019}. We apply the model to reproduce the well sampled SED of the prototypical BL Lac object Mkn 421 obtained during a low state as reported in \citet[][]{Abdo+2011}. The same SED has been modelled with the standard one-zone SSC model by \citet[][]{TavecchioGhisellini2016} and \citet[][]{Tavecchio+2019}, who found a low magnetization and a low radiative efficiency in the emission sites. We show that both these problems may be solved assuming an anisotropic electron population.

The paper is organized as follows. In Section \ref{sec:SED} we develop a phenomenological model of blazar SED assuming an anisotropic electron population. In Section \ref{sec:scenario} we include a self-consistent treatment of the particle cooling into the model. Finally, in Section \ref{sec:discussion} we discuss our results and we conclude. Throughout the paper, the following cosmological  parameters are assumed: $H_0=70{\rm\;km\;s}^{-1}{\rm\; Mpc}^{-1}$, $\Omega_{\rm M}=0.3$, $\Omega_{\Lambda}=0.7$.

\section{Blazar SED from anisotropic electron populations}
\label{sec:SED}

We implement a one-zone SSC model, including the anisotropy of the electron momentum distribution, modifying the model fully described in \citet[][]{MaraschiTavecchio2003}. We assume that the number of electron per unit energy follows a (widely adopted) smoothed broken power law:
\begin{equation}
\label{eq:Ngamma}
N(\gamma)=K\gamma^{-n_1}\left(1+\frac{\gamma}{\gamma_{\rm b}}\right)^{n_1-n_2} \quad {\rm if} \quad \gamma_{\rm min}<\gamma<\gamma_{\rm max}\;.
\end{equation}
The distribution extends from $\gamma_{\rm min}$ to $\gamma_{\rm max}$, and has a break at $\gamma_{\rm b}$. The normalization factor $K$, which has units of ${\rm cm}^{-3}$, controls the total number density of the electrons.

As a phenomenological description of the anisotropy proposed by \citet[][]{SobacchiLyubarsky2019}, we assume that the diffusion of the particle momentum results in a maximum pitch angle that depends on the energy (or Lorentz factor) of the electrons as:
\begin{equation}
\label{eq:alpha}
\theta_{\rm max}\left(\gamma\right) = \frac{\pi}{2}\times
\begin{cases}
1 & {\rm if}\quad \gamma_{\rm min}< \gamma< \gamma_{\rm iso}\\
\left(\gamma/\gamma_{\rm iso}\right)^{-\eta} & {\rm if}\quad \gamma_{\rm iso}<\gamma<\gamma_{\rm max}
\end{cases}
\end{equation}
with $\eta>0$. The velocity of  most energetic electrons, with $\gamma\gg \gamma_{\rm iso}$, is therefore contained within a cone of semi-aperture $\theta_{\rm max}\ll \pi/2$ along magnetic field lines. Instead, the electrons with $\gamma< \gamma_{\rm iso}$ have an isotropic distribution, namely $\theta_{\rm max}=\pi/2$.

We assume the background magnetic field to be tangled, as may happen for example as a result of strong turbulence. An important point is that in the physical conditions characterizing blazars the electron's Larmor radius and the size of the emission region are separated by several orders of magnitude. In these conditions, the electron distribution is {\it locally} (i.e. with respect to the local magnetic field) anisotropic, but {\it globally} (i.e. on the scale of the emission region) isotropic. Hence, the synchrotron emission (which depends on the electron's pitch angle with respect to the local magnetic field) is affected by the local anisotropy of the electron population, while for the IC emission (which is only sensitive to the properties of the electron distribution and the soft radiation field on the scale of the emission region) we can apply the same treatment of the isotropic model used by \citet[][]{MaraschiTavecchio2003}. Also note that, as a consequence of the global isotropy of the electron distribution, the non-thermal emission (i.e. synchrotron and IC) is isotropic in the frame of the emission region.

It is well known \citep[e.g.][]{RybickiLightman1979} that if the electron distribution is isotropic, $\theta_{\rm max}=\pi/2$, and follows a power law, $N(\gamma)\propto \gamma^{-n}$, the synchrotron emissivity displays a power law shape $j(\nu)\propto \nu^{-\alpha}$, with $\alpha=(n-1)/2$. As shown in Appendix \ref{sec:appA}, for an anisotropic distribution following Eq. (\ref{eq:alpha}) the spectrum is described by a power with index $\alpha=(n-1+\eta)/(2-\eta)$ for frequencies above $\nu_{\rm c}(\gamma_{\rm iso})$, where $\nu_{\rm c}(\gamma_{\rm iso})$ is the typical frequency of the photons emitted by the electrons at $\gamma_{\rm iso}$. Therefore, if the electron distribution follows an unbroken power law $N(\gamma)\propto \gamma^{-n}$, the synchrotron spectrum would show a break at the frequency $\nu\sim \nu_{\rm c}(\gamma_{\rm iso})$ with $\Delta\alpha=\eta(n+1)/2(2-\eta)$. Note that $\Delta\alpha$ diverges when $\eta\to 2$. This is due to the fact that $\nu_{\rm c}$ becomes a decreasing function of $\gamma$ when $\eta>2$ (see Appendix \ref{sec:appA}). Hence, one may expect the emissivity to decrease very steeply above $\nu_{\rm c}(\gamma_{\rm iso})$ when $\eta>2$. In the following we focus on the case $\eta<2$.

More formally, the total synchrotron emissivity $j(\nu)$ may be calculated integrating the standard pitch angle-dependent emitted power $P(\nu,\gamma,\theta)$ \citep[e.g.][]{RybickiLightman1979} over the (energy-dependent) range of pitch angles $[0,\theta_{\rm max}(\gamma)]$. We obtain:
\begin{equation}
\label{eq:emiss}
j(\nu)=\frac{\sqrt{3}e^3}{m_ec^2}B \int_{\gamma_{\rm min}}^{\gamma_{\rm max}} N(\gamma) \int_{0}^{\theta_{\rm max}(\gamma)} \sin\theta \, {\cal F} (x) \, {\rm d}\theta \, {\rm d}\gamma\;,
\end{equation}
where the argument of the ${\cal F}$ function \citep[for its explicit form see][]{RybickiLightman1979} is $x=\nu/\nu_{\rm c}(\theta)$, with:
\begin{equation}
\label{eq:nuc}
\nu_{\rm c}(\gamma,\theta)=\frac{3\gamma^2e}{4\pi mc}B\sin\theta\;.
\end{equation}

%%%%%%%%%%%%%%%%%%%%%%%%%%%%%%%%%%
\begin{figure}
\vspace{-1.76truecm}
\hspace{-1truecm}
  \includegraphics[width=.6\textwidth]{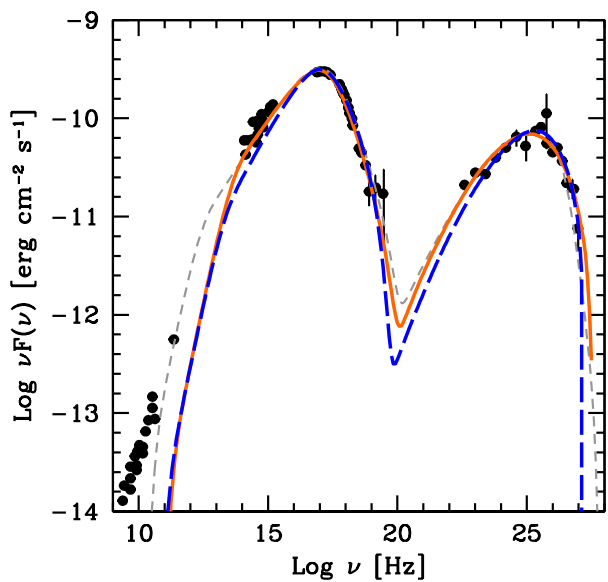} 
  \vspace{-3.4truecm}
  \caption{Spectral energy distribution of Mkn 421 (black filled circles) obtained during the campaign reported in \citet[][]{Abdo+2011} reproduced with the anisotropic electron distribution model assuming a phenomenological electron energy distribution (orange solid: model 1; blue long-dashed: model 2). The parameters are listed in Table \ref{tab:param}. For comparison, the gray dashed line shows the standard isotropic model reported in \citet[][]{TavecchioGhisellini2016}.}
\label{fig:sed421}
  \end{figure}
%%%%%%%%%%%%%%%%%%%%%%%%%%%%%%%%%%

In Fig. \ref{fig:sed421} we report two possible models consistent with the observed SED of the BL Lac object Mkn 421. We assume that the phenomenological electron energy and pitch angle distributions are described by Eqs. \eqref{eq:Ngamma} and \eqref{eq:alpha} respectively (the model parameters are listed in Table \ref{tab:param}). As discussed by \citet[][]{SobacchiLyubarsky2019}, when $\gamma_{\rm iso}<\gamma_{\rm b}$ the synchrotron peak of the SED is produced by the electrons at $\gamma_{\rm iso}$ due to the suppression of the synchrotron emissivity at small pitch angles, while the IC peak is still produced in the Klein-Nishina regime by the electrons at $\gamma_{\rm b}$. We obtain a satisfactory fit using $\gamma_{\rm iso}\sim 10^4$, consistent with the value suggested by \citet[][]{SobacchiLyubarsky2019} for a sample of BL Lacs. The magnetic field is around $1{\rm\; G}$, much larger than what required by the standard isotropic one-zone models, that provides values around $0.1{\rm\; G}$ \citep[e.g.][]{TavecchioGhisellini2016}. The corresponding increased magnetic energy density (by about two orders of magnitude), together with the concomitant reduction of the electron density, allows to reach equipartition keeping all other parameters within the standard range.

\begin{table*}
\centering
\begin{tabular}{ccccccccccccc}
\hline
\hline
Model    & $\gamma _{\rm min}$ &$\gamma _{\rm b}$& $\gamma _{\rm max}$&  $\gamma _{\rm iso}$ &$n_1$&$n_2$ &$\eta$ &$B$ &$K$ &$R$ & $\delta $ & $U_B/U_{\rm e}$\\
$[1]$ & $[2]$  & $[3]$ & $[4]$ & $[5]$ & $[6]$ & $[7]$ & $[8]$  & $[9]$ & $[10]$ & $[11]$ & $[12]$ & $[13]$\\
\hline
1     &$700 $&$ 9\times 10^{4} $&$ 2\times 10^{6} $& $2.8\times 10^4$ & $ 2.2 $&$ 3.2 $ &$0.5$ &$ 1.25 $&$ 7\times 10^{4}$&$ 2.8 $& $15$ & $1.35$ \\      
2     &$500 $&$ 7\times 10^{4} $&$ 10^{6} $& $1.2\times 10^4$ & $ 2 $&$ 3 $ &$0.5$ &$ 1.25 $&$ 1.3\times 10^{4}$&$ 2.65 $& $15$ & $1.30$ \\      
\hline
TG16 &$700 $&$ 2.5\times 10^{5} $&$ 4\times 10^{6} $& $-$ & $ 2.2 $&$ 4.8 $ &$ - $ &$ 0.06 $&$ 3.2\times 10^{3}$&$ 3.6 $& $14$ & $0.053$ \\   
\hline
\hline
\end{tabular}
\vskip 0.4 true cm
\caption{Input model parameters for the models of Mkn 421 in Fig. \ref{fig:sed421} and derived magnetic to electron energy density ratio. For comparison we also report the parameters derived by \citet[][]{TavecchioGhisellini2016} with the standard isotropic model (their model 2).
[1]: model.  
[2], [3], and [4]: minimum, break and maximum electron Lorentz factor. 
[5]: maximum Lorentz factor of electrons with isotropic pitch angle distribution.
[6] and [7]: slope of the electron energy distribution below and above $\gamma_{\rm b}$.
[8]: slope of the electron pitch angle distribution above $\gamma_{\rm iso}$.
[9]: magnetic field in units of G. 
[10]: normalization of the electron distribution in units of cm$^{-3}$. 
[11]: radius of the emission zone in units of $10^{16}{\rm\; cm}$. 
[12]: Doppler factor. 
[13]: ratio between the magnetic and the relativistic electron energy densities.}
\label{tab:param}
\end{table*}

%%%%%%%%%%%%%%%%%%%%%%%%%%%%%%%%%%
\begin{figure}
\vspace{-1.76truecm}
\hspace{-1truecm}
  \includegraphics[width=.6\textwidth]{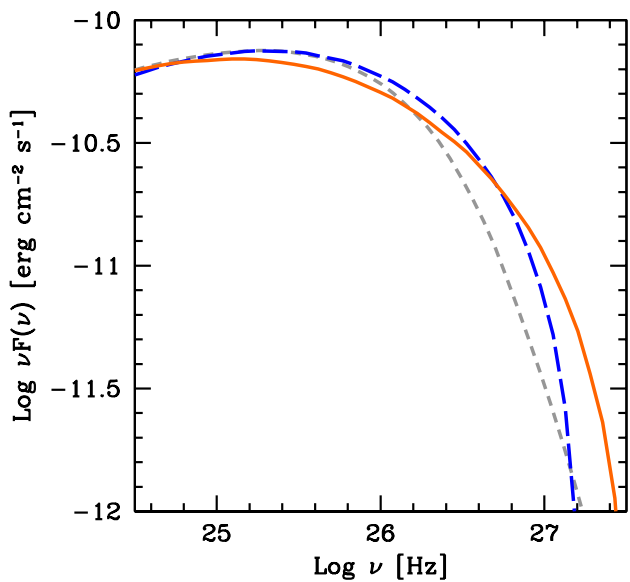} 
  \vspace{-3.4truecm}
  \caption{Zoom of Fig. \ref{fig:sed421} in the VHE range. The anisotropic scenario (blue long-dashed and orange solid lines) predicts SSC spectra harder than the standard model (gray dashed line).}
\label{fig:zoom}
  \end{figure}
%%%%%%%%%%%%%%%%%%%%%%%%%%%%%%%%%%

In the standard isotropic one-zone model the observed slope of the optical-UV and X-ray continua constrain the index of the underlying electron energy distribution in the ranges $n_1\simeq 2-2.2$ and $n_2\simeq 4-4.2$ \citep[e.g.][]{TavecchioGhisellini2016}. While the low-energy index could be interpreted as being produced by cooling (but with the difficulty of the large cooling time mentioned above), the extremely soft high-energy slope is rather difficult to explain. The anisotropic scenario offers a much more coherent scheme. First of all, it allows us to use a harder slope, $n_2\simeq 3$. This, together with the larger radiative losses ensured by the higher magnetic and radiative energy densities, allows one to adopt a scenario in which particle are continuously injected into the emission region with slope $n_{\rm inj}\approx 2$ above a minimum Lorentz factor $\gamma_{\rm inj}\approx \gamma_{\rm b}$.
Radiative losses increase the slope of the equilibrium distribution by $\Delta n=1$ at Lorentz factors $\gamma>\gamma_{\rm inj}$, and produce a $\gamma^{-2}$ tail at $\gamma<\gamma_{\rm inj}$, which results in a suitable energy distribution to reproduce the observed SED. While this scenario is attractive, it is not fully consistent, since it does not take into account that also the synchrotron cooling rate of the high-energy electrons is strongly reduced by the small pitch angles. A self-consistent picture is discussed in the next section.

A last point concerns the IC spectrum. Since in the anisotropic case the underlying electron distribution required to produce the high-energy tail of the synchrotron peak is harder than in the isotropic case, one expects a harder SSC spectrum at high-energy (i.e. after the SSC peak). Indeed, harder spectra are predicted in the VHE by the anisotropic model (Fig. \ref{fig:zoom}). Quite interestingly, in this case, fluxes differences by a factor $2-3$ are expected at few TeV, a prediction that could be easily tested by the upcoming Cherenkov Telescope Array (we have checked that internal absorption is negligible at these energies; see also \citealt{Tavecchio+2019}).

\section{Towards a self-consistent scenario}
\label{sec:scenario}

From the standard synchrotron theory one derives that the cooling rate of electrons with pith angle $\theta$ is:
\begin{equation}
\dot{\gamma}_{\rm s}=\frac{2\sigma_{\rm T}}{m_ec}\gamma^2 \beta^2 \sin^2 \theta \, U_B\;.    
\label{eq:gammadot}
\end{equation}
For an anisotropic distribution one can calculate the average:
\begin{equation}
\langle\dot{\gamma}_{\rm s}\rangle=\frac{2\sigma_{\rm T}}{m_ec}\gamma^2 \beta^2\, U_B \frac{\int_0^{\theta_{\rm max}}\sin^3 \theta \, {\rm d}\theta}{\int_0^{\theta_{\rm max}}\sin \theta \, {\rm d}\theta}=    \frac{4}{3}{\cal A(\gamma)}\frac{\sigma_{\rm T}}{m_ec}\gamma^2 \beta^2 \, U_B\;,
\label{eq:avgammadot}
\end{equation}
where the function ${\cal A(\gamma)}$ is:
\begin{equation}
%{\cal A(\gamma)}=\frac{2-3\mu+\mu^3}{2-2\mu}
{\cal A(\gamma)}=1-\frac{\mu}{2}-\frac{\mu^2}{2}
\label{eq:a}
\end{equation}
with $\mu\equiv\cos\theta_{\rm max}(\gamma)$. The cooling timescale (taking into account the synchrotron losses only) can be defined as:
\begin{equation}
t_{\rm cool,s}(\gamma)\equiv\frac{\gamma}{\langle\dot{\gamma}_{\rm s}\rangle}=\frac{3m_ec}{4{\cal A(\gamma)}\sigma_{\rm T} \, \gamma\, \beta^2 U_B}\;.
\label{eq:tcool}
\end{equation}
Note that when $\gamma<\gamma_{\rm iso}$ (and therefore $\theta_{\rm max}=\pi/2$) the synchrotron cooling time is the same as in the standard isotropic case, while when $\gamma\gg\gamma_{\rm iso}$ (and therefore $\theta_{\rm max}\ll\pi/2$) the synchrotron cooling time is a factor $1/{\cal A(\gamma)}\sim (4/3)\times (1/\theta_{\rm max}^2)$ longer. If the pitch angle distribution is described by Eq. \eqref{eq:alpha}, one finds that $t_{\rm cool,s} \propto \gamma^{2\eta-1}$ when $\gamma\gg\gamma_{\rm iso}$.

%%%%%%%%%%%%%%%%%%%%%%%%%%%%%%%%%%
\begin{figure}
\vspace{-1.76truecm}
\hspace{-1truecm}
  \includegraphics[width=.6\textwidth]{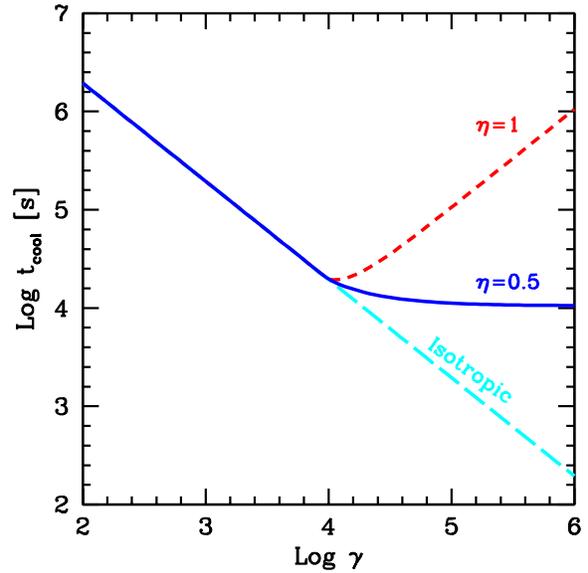} 
  \vspace{-3.4truecm}
  \caption{Synchrotron cooling time as a function of the Lorentz factor for electrons with isotropic (long-dashed cyan) and anisotropic ($\eta=0.5$, solid blue; $\eta=1$, dashed red) momentum distribution. We assume $\gamma_{\rm iso}=10^4$ and $B=2 {\rm\; G}$.}
\label{fig:tcool}
  \end{figure}
%%%%%%%%%%%%%%%%%%%%%%%%%%%%%%%%%%

%%%%%%%%%%%%%%%%%%%%%%%%%%%%%%%%%%
\begin{figure}
\vspace{-1.82truecm}
\hspace{-1.4truecm}
  \includegraphics[width=.65\textwidth]{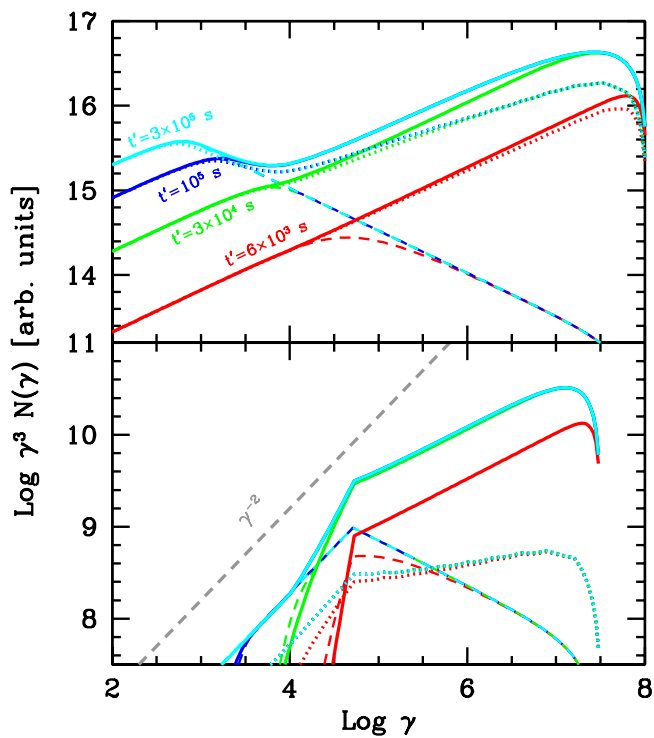} 
  \vspace{-3.truecm}
  \caption{Upper panel: electron energy distribution at different times assuming constant injection $Q(\gamma)\propto \gamma^{-2.5}$ with $\gamma_{\rm inj}=10$ and $\gamma_{\rm max}=10^8$ and the pitch angle distribution Eq. \ref{eq:alpha} with $\eta=0.5$. The dashed line show the distribution in the standard isotropic case. The calculation assumes $\gamma_{\rm iso}=10^4$ and $B=2$ G and are calculated for $t^{\prime}=6\times 10^3$ s (red), $t^{\prime}=3\times 10^4$ s (green), $t^{\prime}=10^5$ s (blue) and $t^{\prime}=3\times 10^5$ s (cyan). The dotted line show the distributions taking into account inverse Compton cooling off photons with a total energy density of $U_{\rm rad}=U_B$ and with the spectral shape of the synchroron component in Fig.\ref{fig:sed421}. Lower panel: as for the upper panel but with injection between $\gamma_{\rm inj}=5\times 10^4$ and $\gamma_{\rm max}=3\times 10^7$.}
\label{fig:ele}
  \end{figure}
%%%%%%%%%%%%%%%%%%%%%%%%%%%%%%%%%%

In Fig. \ref{fig:tcool} we show the synchrotron cooling time as a function of the Lorentz factor for the isotropic case (cyan) and for the anisotropic case with $\eta=0.5$ (blue) and $\eta=1$ (red), and $\gamma_{\rm iso}=10^4$. Clearly, the synchrotron cooling of the electrons is strongly suppressed above $\gamma_{\rm iso}$.
%As discussed below, this has a profound impact on the evolution of anisotropic electron energy distributions. 

An idea of the impact of the anisotropy on the emitted spectrum can be grasped considering the simple case of the steady-state distribution resulting from the balance between continuous injection and radiative losses. In this case, the solution of the continuity equation is given by \cite[e.g.][]{Ghisellini89}:
\begin{equation}
    N(\gamma)=\frac{1}{\dot{\gamma}_{\rm tot}}\int_{\gamma}^{\infty}Q(\gamma^{\prime})d\gamma^{\prime}
\label{eq:stationary}
\end{equation}
where $Q(\gamma)$ is the injection term and $\dot{\gamma}_{\rm tot}$ is the total energy loss rate. In the following we assume an injected power law $Q(\gamma)\propto\gamma^{-n_{\rm inj}}$ for $\gamma_{\rm inj}<\gamma<\gamma_{\rm max}$. For the SSC emission we have $\dot{\gamma}_{\rm tot}=\dot{\gamma}_{\rm s}+\dot{\gamma}_{\rm IC}$.

Note that $\dot{\gamma}_{\rm tot}\propto\gamma^2$ when either (i) $\gamma<\gamma_{\rm iso}$, in which case both both $\dot{\gamma}_{\rm s}$ and $\dot{\gamma}_{\rm IC}$ are proportional to $\gamma^2$, or (ii) $\gamma>\gamma_{\rm iso}$, but the cooling is dominated by the inverse Compton ($\dot{\gamma}_{\rm IC}\gg\dot{\gamma}_{\rm s}$) and the Klein-Nishina effects are not too strong. If $\dot{\gamma}_{\rm tot}\propto\gamma^2$ at all $\gamma$, the solution of Eq. \eqref{eq:stationary} is $N(\gamma)\propto \gamma^{-2}$ at $\gamma<\gamma_{\rm inj}$ and $N(\gamma)\propto \gamma^{-(n_{\rm inj}+1)}$ at $\gamma>\gamma_{\rm inj}$. In the most interesting case $\gamma_{\rm iso}<\gamma_{\rm inj}$, the synchrotron emissivity is characterised by $j(\nu)\propto \nu^{-\alpha}$, where (i) $\alpha=1/2$ if $\nu<\nu_{\rm c}(\gamma_{\rm iso})$; (ii) $\alpha=(1+\eta)/(2-\eta)$ if $\nu_{\rm c}(\gamma_{\rm iso})<\nu<\nu_{\rm c}(\gamma_{\rm inj})$; $\alpha=(n_{\rm inj}+\eta)/(2-\eta)$ if $\nu>\nu_{\rm c}(\gamma_{\rm inj})$. Therefore, the peak of the synchrotron luminosity is produced by the electrons at $\gamma\sim\gamma_{\rm iso}$, as required by the model of \citet{SobacchiLyubarsky2019}.

If instead $\dot{\gamma}_{\rm tot}\propto\gamma^{2-2\eta}$ for $\gamma>\gamma_{\rm iso}$, which happens when the cooling is dominated by the synchrotron ($\dot{\gamma}_{\rm s}\gg\dot{\gamma}_{\rm IC}$), the solution of Eq. \eqref{eq:stationary} is $N(\gamma)\propto \gamma^{-2}$ at $\gamma<\gamma_{\rm iso}$, $N(\gamma)\propto \gamma^{-2+2\eta}$ at $\gamma_{\rm iso}<\gamma<\gamma_{\rm inj}$, and $N(\gamma)\propto \gamma^{-n_{\rm inj}-1+2\eta}$ at $\gamma>\gamma_{\rm inj}$. In this case, the synchrotron emissivity is characterised by $j(\nu)\propto \nu^{-\alpha}$, where (i) $\alpha=1/2$ if $\nu<\nu_{\rm c}(\gamma_{\rm iso})$; (ii) $\alpha=(1-\eta)/(2-\eta)$ if $\nu_{\rm c}(\gamma_{\rm iso})<\nu<\nu_{\rm c}(\gamma_{\rm inj})$; $\alpha=(n_{\rm inj}-\eta)/(2-\eta)$ if $\nu>\nu_{\rm c}(\gamma_{\rm inj})$. It is simple to realise that in this case the peak of the synchrotron luminosity would be produced by the electrons at $\gamma\sim\gamma_{\rm inj}$, and the mechanism invoked by \citet{SobacchiLyubarsky2019} would not operate irrespective of the pitch angle distribution.

To explore the general case of a time-dependent electron energy distribution it is convenient to resort to numerical calculations. We implemented the method described in \citet[][]{ChiabergeGhisellini1999}, who adopted the numerical scheme of \citet[][]{ChangCooper1970} to solve the time-dependent continuity equation including a source term and radiative and adiabatic losses (for simplicity the latter are not considered here). We have used Eq. \eqref{eq:avgammadot} to describe the synchrotron losses for the anisotropic distribution.

As an example of the behavior of the electron distribution, in Fig. \ref{fig:ele} we report the evolution at different times of the electron energy distribution assuming a constant injection $Q(\gamma)\propto \gamma^{-n_{\rm inj}}$ with $n_{\rm inj}=2.5$, and with $\gamma_{\rm iso}= 10^4$ and $B=2{\rm\; G}$. In the upper panel we show the case with $\gamma_{\rm inj}\ll \gamma_{\rm iso}$, while in the lower panel we report the evolution for a case with $\gamma_{\rm inj}>\gamma_{\rm iso}$. The dashed lines show the standard evolution for the isotropic case. 

For the case $\gamma_{\rm inj}\ll \gamma_{\rm iso}$ at different times $t^{\prime}$ the distributions follow the injected spectra up to a Lorentz factor $\gamma_{\rm c}$ for which $t_{\rm cool}(\gamma_{\rm c})=t^{\prime}$. Above this energy the distributions relax and follow the cooled distribution with slope $n_{\rm inj}+1$. For anisotropic distributions, the inefficient cooling caused by the small pitch angles above $\gamma_{\rm iso}$ determines the presence of a hard tail, asymptotically following the uncooled slope $n_{\rm inj}$. For times at which $\gamma_{\rm c}\ll \gamma_{\rm iso}$, above $\gamma_{\rm c}$ the distribution follows the standard one and then rapidly raises to reach the uncooled distribution at $\gamma\gg \gamma_{\rm iso}$. Of course, when the synchrotron cooling becomes very small, the IC losses can be relevant, even if scatterings occur deeply in the KN regime. This is shown by the dotted curves which include both synchrotron and IC losses (note that the IC scattering is not efficient to produce pitch angle diffusion, see the discussion in Appendix \ref{sec:appB}). The IC cooling is calculated assuming the photon energy density associated to the synchrotron component in Fig. \ref{fig:sed421}. 

For the case $\gamma_{\rm inj}>\gamma_{\rm iso}$ (lower panel), below $\gamma_{\rm inj}$ the cooling produces a tail extending to the cooling Lorentz factor defined by $\gamma_{\rm c}$ for which $t_{\rm cool}(\gamma_{\rm c})=t^{\prime}$. When the IC losses are considered, the slope of this cooling tail follows the standard $\gamma^{-2}$ slope. However, if IC losses are negligible (solid lines), the suppression of the synchrotron cooling determines a slope in the interval $\gamma_{\rm iso}<\gamma<\gamma_{\rm inj}$ harder than $-2$ (see discussion above).

\begin{figure}
\vspace{-1.76truecm}
\hspace{-1truecm}
  \includegraphics[width=.6\textwidth]{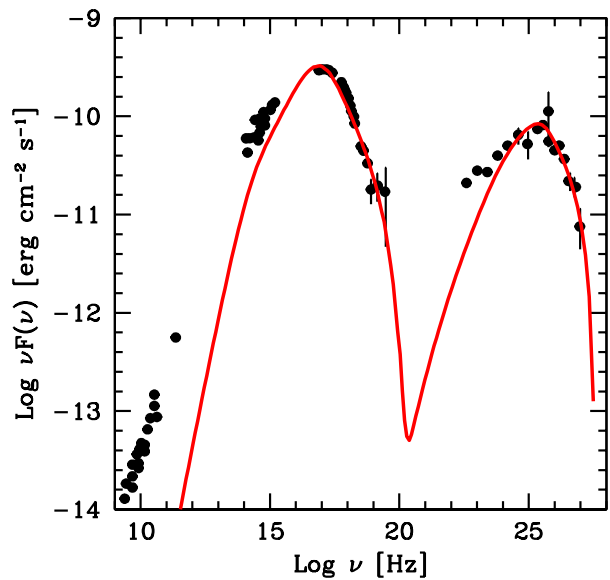} 
  \vspace{-3.4truecm}
  \caption{SED of Mkn 421 reproduced with the anisotropic model assuming a self-consistent electron distribution. See text for details.}
\label{fig:sedself}
  \end{figure}

We tried to reproduce the SED of Mkn 421 by using a self-consistent electron energy distribution, derived by considering the time-evolved distribution at a time $t^{\prime}_{\rm inj}=2R/c$. The injection term is described by a power law between $\gamma_{\rm inj}=6\times 10^4$ and $\gamma_{\rm max}=3\times 10^6$ with slope $n_{\rm inj}=2.75$. Below $\gamma_{\rm inj}$ a cooled tail $\propto \gamma^{-2}$ develops down to a Lorentz factor for which the cooling time equal the injection time. A good agreement with the data can be found if we further assume $B=0.9{\rm\; G}$, $\delta=10.5$, $R=4\times 10^{15}{\rm\; cm}$, $\gamma_{\rm iso}=1.5\times 10^4$, $\eta=0.5$. The result is shown in Fig. \ref{fig:sedself}. It should be remarked that in this case the IC losses are relevant for energies around $\gamma_{\rm iso}$ and this allows the formation of the standard $\gamma^{-2}$ tail. In this situation the pitch angle suppression of the emission can operate and the SED can be reproduced in equipartition condition (we derive $U_B/U_{\rm e}=2.1$). 
%As already remarked above, the presence of an efficient (i.e. not in full KN regime) IC cooling for the electrons with Lorentz factor between $\gamma_{\rm iso}$ and $\gamma_{\rm inj}$ is mandatory for the mechanism to work.

%%%%%%%%%%%%%%%%%%%%%%%%%%%%%%%%%%
%%%%%%%%%%%%%%%%%%%%%%%%%%%%%%%%%%

\section{Discussion}
\label{sec:discussion}

We have developed a model for the SED of the high-energy emitting BL Lacs, implementing the idea proposed by \cite{SobacchiLyubarsky2019} that the momentum distribution of the most energetic electrons is anisotropic. The reduced pitch angle of the electrons above a certain Lorentz factor $\gamma_{\rm iso}$ implies a severe suppression of their synchrotron emission. If $\gamma_{\rm iso}< \gamma_{\rm b}$, where $\gamma_{\rm b}$ is the Lorentz factor of the electrons emitting at the synchrotron and the IC peaks of the SED in the standard SSC model, the anisotropy has an important effect on the modelling of the observed emission. We have shown that the representative SED of the BL Lac object Mkn 421 can be reproduced under equipartition between the magnetic field and the relativistic electrons, contrary to what happens with the commonly adopted isotropic scenario. We have also explored a self-consistent scheme in which the electrons follow an energy distribution resulting from the balance between injection and cooling (synchrotron and IC). In this case we have remarked that the mechanism proposed by \citet{SobacchiLyubarsky2019} can operate only in the presence of an effective IC cooling of the electrons with $\gamma\gtrsim \gamma_{\rm iso}$.

Apart from the stronger magnetic field, the values of the parameters derived for the jet of Mkn 421 in the anisotropic scenario are similar the usual ones \cite[e.g][]{Tavecchio2010}. It is however worth noting that the Doppler factor can be kept to somewhat smaller values than those required by the standard model \citep[see e.g.][]{TavecchioGhisellini2016}. This implies a significantly larger radiation energy density in the emission sites,\footnote{It is worth remarking that the radiation energy density is proportional to $\delta^{-4}$, where $\delta$ is the Doppler factor.} with consequences for the transparency of the source at TeV energies and a potential impact on the multimessenger role of highly-peaked BL Lacs \citep{Tavecchio+2019}.  

From the physical point of view, we remark that the scenario sketched in this paper is not fully self-consistent. In particular, our choice for the pitch angle distribution is purely phenomenological. The detailed description of the anisotropy as a function of the energy has a strong impact on the derived spectra and it would therefore be important to have a physically-motivated model for it. The mechanism proposed by \citet{SobacchiLyubarsky2019} requires (i) the electrons to be accelerated in the direction of the magnetic field; (ii) a significant proton component to be present, thus providing the isotropization of the least energetic electrons. Regarding the item (i), promising candidates are reconnection with a strong guide field and/or dissipation of relativistic turbulence. Interestingly, recent kinetic simulations of decaying turbulence have suggested that the particles at $\gamma\sim\gamma_{\rm inj}$ are primarily accelerated by an electric field aligned with the local magnetic field \citep[e.g.][]{ComissoSironi2018}. Unfortunately, these simulations did not model radiative cooling and were limited to the case of a pair plasma. Regarding the item (ii), there are compelling arguments supporting the presence of a (cold) proton component in FSRQ \citep[e.g.][]{SikoraMadejski2000, GhiselliniTavecchio2010, Ghisellini2014}, but for BL Lacs the situation is less clear. The potential association of the BL Lac object TXS 0506+056 with high-energy neutrinos detected by IceCube \citep{TXS18} (thought to be produced through photomeson interactions involving relativistic protons and radiation) supports the existence of an important barionic component also in BL Lac jets.

\section*{Acknowledgments}
We thank the referee for useful comments.
We thank G. Ghisellini and Y. Lyubarsky for useful discussions. FT acknowledges contribution from the grant INAF CTA--SKA ``Probing particle acceleration and $\gamma$-ray propagation with CTA and its precursors'' and the INAF Main Stream project ``High-energy extragalactic astrophysics: toward the Cherenkov Telescope Array''. ES acknowledges contribution from the Israeli Science Foundation (grant 719/14) and from the German Israeli Foundation for Scientific Research and Development (grant I-1362-303.7/2016).

%%%%%%%%%%%%%%%%%%%%%%%%%%%%%%%%%%%%%%%%%
\def\aap{A\&A}\def\aj{AJ}\def\apj{ApJ}\def\apjl{ApJ}\def\mnras{MNRAS}\def\prl{Physical Review Letters}
\def\araa{ARA\&A}\def\physrep{PhR}\def\sovast{Sov. Astron.}\def\nar{NewAR}
\def\aapr{Astronomy \& Astrophysics Review}\def\apjs{ApJS}\def\nat{Nature}\def\na{New Astron.}
\def\prd{Phys. Rev. D}\def\pre{Phys. Rev. E}\def\pasp{PASP}\def\ssr{Space Sci. Rev.}
\bibliographystyle{mn2e}
\bibliography{paper.bib}

\appendix

\section{Synchrotron spectrum from anisotropic electrons}
\label{sec:appA}

We calculate the synchrotron emissivity for frequencies above that emitted by the electrons at $\gamma_{\rm iso}$. From Eq. \eqref{eq:nuc} one may calculate the typical synchrotron frequency emitted by the electrons at $\gamma$ as
\begin{equation}
\label{eq:A1}
\nu\sim\nu_{\rm c}\left(\theta_{\rm max}\right)\sim \frac{3\gamma^2e}{4\pi mc}B\theta_{\rm max}\propto \gamma^{2-\eta}\;,
\end{equation}
where we have used our assumption that $\theta_{\rm max}\propto\gamma^{-\eta}$, and the fact that $\theta_{\rm max}\ll\pi/2$ when $\gamma\gg\gamma_{\rm iso}$ (see Eq. \ref{eq:alpha}). Note that $\nu_{\rm c}$ becomes a decreasing function of $\gamma$ when $\eta>2$.

The dependence of the total power $\nu j(\nu)$ emitted at the frequency $\nu$ on the Lorentz factor $\gamma$ can be estimated as
\begin{equation}
\label{eq:A2}
\nu j(\nu) \propto \gamma^3 N(\gamma) B^2\sin^2\theta_{\rm max}\propto\gamma^{3-n-2\eta}\;,
\end{equation}
where we have assumed that the electron distribution follows a power law $N(\gamma)\propto\gamma^{-n}$.

Inverting Eq. \eqref{eq:A1} we find that $\gamma\propto\nu^{1/(2-\eta)}$. Substituting into Eq. \eqref{eq:A2}, we finally get
\begin{equation}
\nu j(\nu) \propto \nu^{(3-n-2\eta)/(2-\eta)}\;,
\end{equation}
or equivalently
\begin{equation}
j(\nu)\propto \nu^{(1-n-\eta)/(2-\eta)}\;.
\end{equation}
Note that if the pitch angle does not depend on the energy ($\eta=0$) one recovers the familiar result that $j(\nu)\propto\nu^{(1-n)/2}$.

\section{Pitch angle diffusion due to inverse Compton scattering}
\label{sec:appB}

Let us consider a particle of Lorentz factor $\gamma\gg 1$ that is inverse Compton scattering $\dot{N}$ photons of energy $\varepsilon$ per unit time. After each scattering, a photon of energy $\gamma^2 \varepsilon$ is emitted within an angle of $1/\gamma$ with respect of the direction of the particle motion. Hence, the particle looses an amount of energy $\Delta E \sim \gamma^2 \varepsilon$, and the cooling time (due to IC scattering only) can be calculated as
\begin{equation}
t_{\rm cool, IC} \sim \frac{1}{\dot{N}}\frac{mc^2}{\gamma \varepsilon}\;.
\end{equation}
The particle pitch angle instead varies in a diffusive way. At each scattering, the momentum transverse to the direction of motion that is transferred to the particle is $\Delta p\sim (1/\gamma)\times (\gamma^2\varepsilon/c)\sim \gamma\varepsilon/c$. Hence, the pitch angle varies by $\Delta\theta\sim\Delta p/\gamma mc\sim \varepsilon/mc^2$. Over a cooling time, the pitch angle diffuses by
\begin{equation}
\theta\sim\sqrt{\dot{N}t_{\rm cool, IC}(\Delta\theta)^2}\sim \sqrt{\frac{\varepsilon}{\gamma mc^2}}\;.
\end{equation}
Since the IC scattering is efficient only if $\varepsilon\lesssim mc^2/\gamma$ (otherwise the scattering occurs in the Klein-Nishina regime, which strongly suppresses the cross section), one finds that $\theta\lesssim 1/\gamma$. Hence, the pitch angle diffusion due to IC scattering may be neglected.

\end{document}